\documentclass[a4paper,useAMS,usenatbib,fleqn]{mn2e}
\usepackage{graphics,graphicx,times}
\usepackage{amsmath}
\usepackage{amssymb}
\usepackage{aas_macros}
\usepackage[usenames,dvipsnames]{xcolor}

\newcommand{\sigmasfr}{\dot \Sigma_{\rm \star}}

\newcommand{\qii}{\hbox{$Q_{\rm H II}$}}

\title[]{The duration of reionization constrains the ionizing sources}
\author[Mahavir Sharma et al.]
{Mahavir Sharma\thanks{mahavir.sharma@durham.ac.uk}, Tom Theuns, Carlos Frenk\\
Institute for Computational Cosmology, Department of Physics, University of Durham, South Road, Durham, DH1 3LE, UK\\
}

\begin{document}

\date{Submitted ---------- ; Accepted ----------; In original form ----------}


\maketitle
\begin{abstract}
  We investigate how the nature of the galaxies that reionized the Universe affects the duration of reionization. We contrast two sets of models: one in which galaxies on the faint side of the luminosity function dominate the ionizing emissivity, and a second
	in which the galaxies on the bright side of the luminosity function dominate. The faint-end of the luminosity function evolves slowly, therefore the transition from mostly neutral to mostly ionized state takes a much longer time in the first set of models compared to the second. Existing observational constraints on the duration of this transition are relatively weak, but taken at face value prefer the model in which galaxies on the bright side play a major role. Measurements of the kinetic Sunyaev Zeldovich effect in the cosmic microwave background
	from the epoch of reionization also point in the same direction.

\end{abstract}

\section{Introduction}
Recent measurements of the Thomson optical depth to the surface of last scattering by the {\sc planck} satellite \citep{Planck16} indicate that the reionization of hydrogen in the Universe completed between redshifts $\approx6\hbox{--}9$. The nature of the sources of ionizing photons is currently unknown, with population-III stars \citep[e.g.][]{Sokasian04,Loeb01}; the first galaxies \citep[e.g.][]{Haardt12,Robertson13} and quasars \citep[e.g.][]{Haardt15,Mitra16} plausible candidates. Early galaxies are currently the most popular \citep{Robertson15, Bouwens15, Sharma15}.

Models of reionization that invoke galaxies as sources start by fitting the UV-luminosity functions observed at high redshift \citep[e.g.][]{Bouwens15}. The corresponding ionizing emissivity is calculated by integrating the fit weighted by a factor known as the \lq escape fraction\rq\ of ionizing photons to account for absorption in the galaxy by gas and dust. A number of studies \citep[e.g.][]{Robertson13} have found that, in order to generate the ionizing emissivity required for reionization, either a relatively high escape fraction has to be assumed (of order 20 per cent or more), or that the luminosity function has to be extrapolated to extremely low luminosities. The latter effectively implies assuming that reionization is driven by a large population of galaxies yet to be discovered. \citet{Robertson15} argue that with a constant escape fraction of 20 percent and by extrapolating the luminosity function to a UV (1500~\AA) magnitude of $-13$, the current {\sc planck} measurements can be easily accounted for \citep[see also][]{Bouwens15,Mitra15}. However, the assumption of a constant escape fraction in such models effectively implies a major contribution of ionizing photons from faint, as of yet undetected, galaxies.

There is little theoretical or observational motivation for assuming that the escape fraction is the same for all galaxies.
	For example the escape fraction of the Milky Way is thought to be much less than 20 per cent, whereas the Lyman Break Analogs (LBAs) observed by \citep[e.g.][]{Borthakur14,Izotov16} have an escape fraction of $\gtrapprox 10$~per cent. In addition, there is a strong indication that the mean escape fraction evolves with redshift \citep[e.g.][]{Haardt12,Khaire15,Faisst16,Sharma17} in order to explain the shape of the photoionization background inferred from the Lyman-$\alpha$ forest. How the escape fraction depends on other galaxy properties is clearly key in understanding reionization.

Radiation hydrodynamic simulations that compute the escape fraction of simulated galaxies find that the escape fraction is higher when a galaxy is going through a bursty phase of vigorous star formation \citep[e.g.][]{Wise09,Wise14,Kimm14,Ma16,Trebitsch17}.
	However they disagree on the details, for example on the exact value of the escape fraction, plausibly because this depends sensitively on the gas distribution on very small scales which are challenging to model. For example \cite{Wise09} report that the escape fraction is lower for lower-mass faint galaxies whereas \cite{Paardekooper11} find the opposite trend \citep[see also][]{Wise14,Ma16}. 

In \cite{Sharma15,Sharma17} we presented a phenomenological model for the dependence of the escape fraction on star formation activity in galaxies. In this model, the escape fraction is linked to the star formation rate surface density ($\sigmasfr$), since that is the quantity that governs whether the feedback from star formation is able to drive outflows creating channels through which ionizing photons can escape. In such a model, the fainter galaxies that are yet undetected ($M_{1500}\gtrsim  -16$) make a limited contribution to the total ionizing emissivity, even for a steep faint end slope, because galaxies on the bright side of the luminosity function (already detected in the Hubble deep field) have high escape fractions due to their high star formation activity. These brighter galaxies account for about half of the ionizing emissivity required for reionization \citep{Sharma15}. Interestingly, some recent observations support this viewpoint as they find a dependence of escape fraction on the surface density of star formation for some nearby starburst galaxies that have a high $\sigmasfr$ \citep{Borthakur14,Izotov16}. The simulations by \cite{Trebitsch17} also confirm that supernovae feedback controls the escape fraction of ionizing photons. 

In addition to the redshift at which reionization occurred, the duration of the epoch of reionization is a key parameter, as we will show in this {\em Letter}. The evolution of quantities such as the volume filling factor of ionized hydrogen and the global HI 21~cm brightness temperature is sensitive to the rate at which the emissivity builds up, which in turn depends on the evolution of the luminosity function of galaxies. The luminosity function evolves much more rapidly at the bright end than at the faint end between redshift $10$ and $6$ \citep{Bouwens14}. Therefore, the speed at which reionization progresses may indicate whether galaxies on the bright side of the luminosity function provided a larger share of photons than the ones on the fainter side.

In this {\em Letter} we investigate the temporal evolution of the ionized fraction based on two sets of analytical models; one in which faint galaxies dominate as sources of ionizing photons and another in which the bright galaxies dominate. An estimate of the duration of reionization can be obtained by studying the evolution of the ionized fraction. We calculate the evolution of ionized fraction and compare with  existing constraints obtained from various observations:  Ly$\alpha$ dark gap statistics \citep{McGreer15}), the IGM damping wings in a $z=7$ quasar \citep{Mortlock11}, the damping wing in a Gamma-ray burst \citep{Totani14}, galaxy clustering \citep{Mcquinn07}, Ly$\alpha$ emitters \citep{Ota08,Ouchi10} and the Ly$\alpha$ emission statistics of galaxies \citep{Caruana12,Tilvi14,Schenker14}.

The redshift and duration of reionization can also be constrained by the measurement of the brightness temperature of the global HI 21~cm line signal \citep[e.g.][]{Monslave17}. Measurements of this global signal is different from the measurements in its fluctuations targeted by {\sc lofar} \citep{van-Haarlem13} or {\sc ska} \citep{Pritchard15}. The global signal can instead be measured using single-dish interferometric experiments, many of which have been proposed in the past two decades \citep[e.g.][]{Bowman08,Singh17}. A recent study that uses the data obtained from the {\sc edges} experiment \citep{Bowman08} reported preliminary constraints on the redshift and set a {\em lower} limit on the duration of reionization \citep{Monslave17}. On the other hand, the measured amplitude of the patchiness in the kinematic Sunyaev Zeldovich (kSZ) effect twoards the surface of last scattering \citealt[e.g.][]{Zahn12}) can be used to set an {\em upper} limit on the duration of reionization \citep{Zahn12, George15}. We compare our models with these recent observational constraints.

This {\em Letter} is organised as follows. In section 2 we describe our models and assumptions. In section 3, we present our results on the evolution of the ionized/neutral fraction and compare it with the observations. We conclude our findings in section 4.  
\section{Model description}

\begin{table}
	\centering
	\caption{The functional form of the escape fraction ($f_{\rm esc}$) used in three of our fiducial models. The variable, $f_{\rm esc,\star}$, represents the maximum allowed value of escape fraction for models F10, F14 and B16, and, it corresponds to the value of escape fraction at redshift 7 for models F10z and F14z. $\mathcal{H}=\mathcal{H}(M_{\rm cut}  -  M_{\rm 1500})$ is the Heaviside step function where $M_{\rm cut}$ is the magnitude at which the escape fraction steps up or down.}

	\label{tab:example_table}
	\begin{tabular}{lccr} 
		\hline
		Model & $f_{\rm esc}$ & $M_{\rm cut}$ & $M_{\rm low}$ \\
		\hline
		F10 & $f_{\rm esc,\star} (1-\mathcal{H})$ & -10 & -6\\
		F10z & min[$f_{\rm esc,\star} (1-\mathcal{H}) (1+z)/8$, 1] & -10 & -6\\
		F14 & $f_{\rm esc,\star} (1-\mathcal{H})$ & -14 & -6\\
		F14z & min[$f_{\rm esc,\star} (1-\mathcal{H}) (1+z)/8, 1]$ & -14 & -6\\
		B16 & $f_{\rm esc,\star} \mathcal{H}$ & -16 & ----\\
		\hline
	\end{tabular}
	\label{tab_mod}
\end{table}
 The luminosity functions of galaxies is usually fit with a Schechter function (Fig.~\ref{fig_Lfunc}). Such a function combines a power law at the faint end with an exponential cut off at the bright end. Such fits serve as an input for an ab initio calculation of reionization that begins with an estimate of the total number of UV photons produced at any given redshift, which then can be converted into the number of ionizing photons by a conversion factor calculated from a population synthesis model \citep[e.g.][]{Schaerer03}. The ionizing emissivity is then,
\begin{equation}
\dot n_{\rm \gamma, esc} (z) = \int_{L_{\rm low}}~f_{\rm esc}~f_{912}~L_{1500}~\phi(L_{1500},z)~dL_{1500}\,,
\label{eq:emis}
\end{equation}
where $\phi(L_{1500},z)$ is the luminosity function  (\citealt{Bouwens14}); $L_{1500}$ is the luminosity at $1500$~\AA; $L_{\rm low}$ is the faint end limit of the luminosity that corresponds to a UV-magnitude, $M_{\rm low}$; $f_{\rm 912}$ is the conversion factor from $L_{1500}$ to the luminosity at the Lyman limit, $L_{912}$ \citep{Schaerer03}; and $f_{\rm esc}${\footnote{$f_{\rm esc}$ is a combination of two factors that account for absorption by dust and absorption by gas. In this study we assume that the dust has a minor effect at redshift $z>6$, as stated in \citet{Bouwens14}.} is the escape fraction of ionizing photons from a galaxy, which may depend on the luminosity (or other properties) of the galaxy. Here, we explore the consequences of this dependence for the history of reionization.

We consider five fiducial models for $f_{\rm esc}$, summarised in Table~\ref{tab_mod}. In models F10 and F14, galaxies fainter than $M_{\rm 1500}=-10$ and $-14$, respectively, have a constant escape fraction, $f_{\rm esc,\star}=0.2$, whereas those brighter than these limits have $f_{\rm esc}=0$. Such a choice is inspired by models in which faint galaxies (below the {\em Hubble Ultra Deep} field detection limit) are the main	drivers of reionization ({\em e.g.} \citealt{Bouwens12, Robertson13}). In model B16 in contrast, galaxies {\em brighter} than $M_{\rm 1500}=-16$ have a non-zero escape fraction, whereas those fainter than this limit have $f_{\rm esc}=0$. We chose these contrasting models to examine whether the nature of the sources affects the reionization history; they are illustrated in Fig.~\ref{fig_Lfunc}. 

\begin{figure}
\centering
\includegraphics[width=\linewidth]{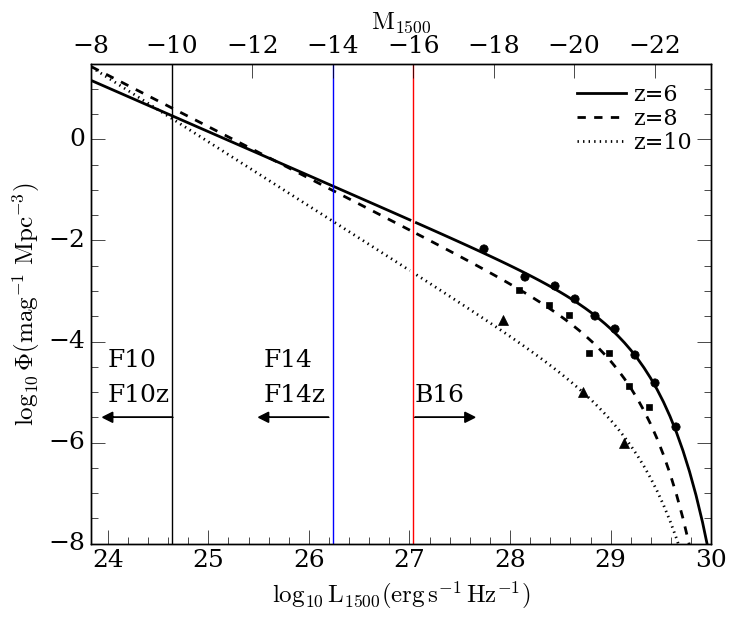}
\caption{Fits to the UV luminosity functions from \citet{Bouwens14} used in this study for redshift $z=6$ (solid), $8$ (dashed), $10$ (dotted); the actual measurements shown as circles, squares and triangles for redshifts 6, 8 and 10 respectively. The luminosity, $L_{1500}$, on the bottom x-axis and the corresponding magnitude, $M_{1500}$, is given on the top-axis. Vertical lines illustrate our fiducial models (Table~\ref{tab_mod}) with arrows indicating the portion of the luminosity function with non-zero escape fraction: black line for models F10 and F10z, blue line for models F14 and F14z, and red line for model B16 (see Table~\ref{tab_mod}).}
\label{fig_Lfunc}
\end{figure}

The escape fraction may also depend on redshift, in addition to luminosity. In their simulations, \cite{Wise14} reported a decrease in escape fraction with increasing virial mass thereby finding that the escape fractions are higher for fainter galaxies; in fact, significant ($\gtrsim 10$~percent) escape fractions are expected only in the faintest galaxies ($M_{\rm 1500}>-10$). Moreover, in these simulations, the escape fractions of galaxies are found to increase with increasing redshift \citep[see also][]{Yajima11, Ma16}. To mimic these results, we consider two additional models, F10z and F14z, that are similar to F10 and F14 except that the constant escape fraction is replaced with one that evolves with redshift (see Table.~\ref{tab_mod}). This gives us a grand total of five models.

We substitute $f_{\rm esc}$ of these models in Eq.~(\ref{eq:emis}) and calculate the emissivity; the results are plotted in Fig.~\ref{fig_E}: models F10, F10z are plotted as black line, F14, F14z as blue lines and B16 as a red line. We have extrapolated the emissivity at redshifts greater than 10, where there are currently no data, and plot the results as a dashed line. 
 For models in which the galaxies on the faint side dominate, the emissivity is approximately constant from redshift 8 to 10. We have assumed it to remain constant up to $z=15$, after which it falls rapidly to $z=25$. This extrapolation has little effect on our results, basically because the elapsed time is small. 
 
Our motivation to distinguish between {\em faint} and {\em bright} galaxies stems from the dramatic difference in the rate of evolution of the number density of low mass and massive dark matter haloes that host such galaxies; the demarcation mass corresponds to the \lq knee\rq\ of the Press-Schechter mass function. For example using Fig.~8 in \cite{Reed07}, the number density of haloes with mass $10^9$~M$_\odot$ increases by a factor of 5 between $z=10$ and $z=8$, but at $10^{11}$~M$_\odot$ the increase is $\sim 100$. Of course we do not know the halo mass function for the observed galaxies; we chose $M_{1500}=-16$, approximately two magnitudes fainter than the \lq knee\rq~ in the observed Schechter luminosity function, to distinguish between \lq faint\rq\ galaxies whose emissivity evolves slowly and \lq bright\rq\ galaxies that evolve rapidly. In fact, even if the emissivity for faint galaxies dominated models were to decrease beyond redshift 10, which we think is contrived and not well motivated, such a decrease is unlikely to be as rapid as the nearly exponential rate associated with model B16. We show below that any rate much less rapid than that for model B16 is disfavoured by current data.

 Cumulatively down to redshift 6, all of our models produce approximately the same number of photons, yielding approximately the same redshift of reionization (see Figure~\ref{fig_E}, top panel). 
The emissivity for model B16 shows a steeper increase than in the other four models, a consequence of the fact that the bright side of the luminosity function builds up rapidly with decreasing redshift, whereas the faint side was in existence at an earlier redshift and evolves minimally from redshift 10 to 6. We study the consequences of this on the history of reionization in the next section.

\section{Results}
\begin{figure}
\centering
\includegraphics[width=\linewidth]{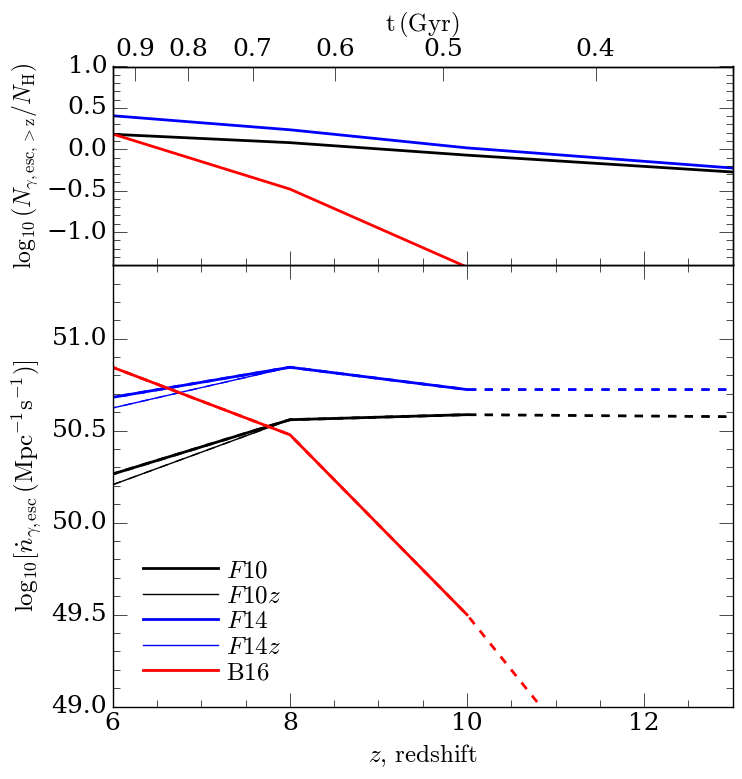}
\caption{Bottom panel: The evolution of the emissivity of ionizing photons as a function of redshift for our fiducial models, F10 (black), F14 (blue) and B16 (red). The models F10z and F14z are shown as the corresponding thin black and blue lines. The dashed portion of the curves shows the extrapolation that we have adopted at redshifs greater than 10. For models in which faint galaxies dominate (shown by the black and blue curves), the emissivity is assumed to stay constant up to redshift 15 \citep[e.g.][]{Wise14,Ma16} followed by a decrease of more than 3 orders of magnitudes to redshift 25. Top panel: The cumulative number of photons per hydrogen atom that escape from galaxies up to a given redshift for models F10 (black), F14 (blue) and B16 (red).
}
\label{fig_E}
\end{figure}
\begin{figure}
\centering
\includegraphics[width=0.98\linewidth]{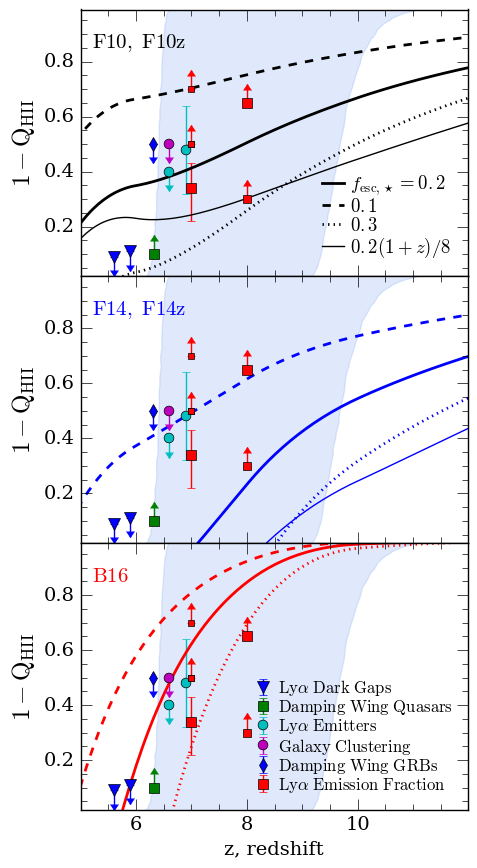}
\caption{ Top panel: the filling factor of neutral hydrogen, $1-Q_{\rm HII}$, for  model F10, for a constant escape fraction, $f_{\rm esc,\star}=0.1$ (dashed), 0.2 (solid) and 0.3 (dotted curve). The model F10z, for which the escape fraction evolves with redshift, is shown using a thin black curve. Middle panel: same as the top panel but for models F14 and F14z. Bottom panel: same as the top panel but for model B16 in which the galaxies with $M_{1500}>-16$ contribute to the emissivity. For comparison we also show the observational estimates, from Ly$\alpha$ dark gap statistics (blue triangles, \citealt{McGreer15}), the damping wings in  a $z=7$ quasar (green square, \citealt{Mortlock11}), the damping wing in Gamma ray burst (black diamond, \citealt{Totani14}), galaxy clustering (magenta circle, \citealt{Mcquinn07}), Ly$\alpha$ emitters (cyan circles, \citealt{Ota08,Ouchi10}) and the Ly$\alpha$ emission statistics of galaxies {\protect\citep{Caruana12,Tilvi14,Schenker14}}. The light blue-shaded region is inferred from {\sc planck} \citep{Planck16} at 95~percent confidence. }
\label{fig_Q}
\end{figure}
\subsection{Evolution of the ionized fraction}
The mean volume filling factor of ionized regions, \qii, quantifies the history of cosmic reionization. The equation describing its evolution features a source and a sink term,
\begin{equation}
\dot Q_{\rm HII}={\dot n_{\rm \gamma, esc}\over \langle n_{\rm H}\rangle} - 1.08\,\alpha_{\rm B}\,{\cal C}\langle n_{\rm H}\rangle Q_{\rm HII}, 
\label{eq:Q2}
\end{equation}
with the first term the rate at which \ion{H}{II} is produced through photo-ionization, and the second term the rate at which it is lost due to recombinations. Here, $\dot n_{\rm \gamma,esc}$ is the rate at which ionizing photons are emitted per unit proper volume (see Eqn.~\ref{eq:emis}), $\langle n_{\rm H}(z)\rangle$ is the mean proper hydrogen number density; the factor 1.08  is to account for the reionization of \ion{He}{i} to \ion{He}{ii}; $\alpha_{\rm B}$, the recombination coefficient; and ${\cal C}\equiv \langle n_{\rm   H}^2\rangle/\langle n_{\rm H}\rangle^2$ is the clumping factor from simulations \citep[e.g.][]{Pawlik09}. A radiative transfer calculation is required to account for the effects of Lyman Limit Systems (LLSs), or to compute spatial variations in \qii\ \citep[e.g.][]{Shukla16}. Nevertheless, the widely used Eqn.~(\ref{eq:Q2}) gives a reasonable description of the global reionization history \citep{Haardt12, Robertson15, Bouwens15, Mitra15, Gnedin16} (but see \citealt{Madau17} for an improvement on this equation).

We integrate Eq.~(\ref{eq:Q2}) for our five models and plot the result in Figure~\ref{fig_Q}. As expected, in model B16, \qii\ transitions rapidly from mostly neutral to mostly ionized, because the emissivity changes rapidly with redshift. In contrast in the other models, the slowly evolving luminosity function results in a gentle build-up of the emissivity, and consequently the transition in \qii\ from neutral to ionized takes much longer.

How do these model histories compare to observational data? In Figure~\ref{fig_Q}, we over plot current observational constraints as data points with error bars. The data points all suggest a decrease in $1-Q_{\rm HII}$ around $z=7$, consistent with the 
	{\sc planck} limits on reionzation (light blue region). The transition is relatively rapid, from 80~percent neutral at redshift 8, to almost fully ionized at redshift 6.  This is the trend seen in model B16, and is caused by the rapid build-up of the emissivity as the bright end of the luminosity function evolves rapidly. The transition is much more gentle in the other models. However, the current data is not very constraining, and better constraints are needed to conclusively rule out a model such as F10 or F14 in which the galaxies on the faint side of luminosity function dominate the ionizing emissivity. 

We further quantify the progress of reionization in the models that we have presented by two parameters: the redshift of reionization, $z_{\rm re}$ (defined as the epoch where \qii=0.5) and, the duration of reionization following
	\citet{Planck16}}, 	$\Delta z = z_{\rm beg} - z_{\rm end}$, where $z_{\rm beg}$ and $z_{\rm end}$ are the redshifts at which $Q_{\rm HII}=0.1$ and $Q_{\rm HII}=0.99$, respectively. We calculate $z_{\rm re}$ and $\Delta z$ for our models and plot them in Figure~\ref{fig_dz}. Models F10, F14, F10z and F14z, in which galaxies on the faint side of the luminosity function dominate, all yield $\Delta z\gtrapprox 8$. Model B16, in which galaxies on the bright side dominate, has a much shorter duration, $\Delta z \lessapprox 2$. 
\begin{figure}
\centering
\includegraphics[width=\linewidth]{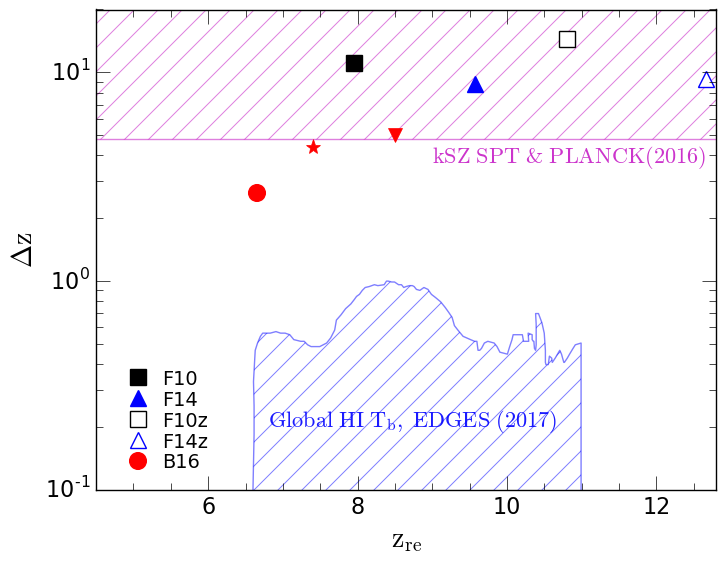}
\caption{The duration of reionization ($\Delta z$) as a function of the redshift of reionization ($z_{\rm re}$) for model F10 (Filled black square), F10z (open black square), F14 (filled blue triangle), F14z (open blue triangle) and B16 (filled red circle); see Table~\ref{tab_mod} for details of the models. For comparison, we also show $z_{\rm re}$ and $\Delta z$ from the models of \citet{Robertson15} (red star), and \citet{Mitra15} (red triangle).   The rejection zone proposed by \citet{Planck16} based on measurements of the kSZ effect and electron scattering optical depth is depicted as a magenta hatched zone.  The blue hatched zone represents the region excluded by using the measured global HI 21~cm brightness temperature by \citet{Monslave17}; they use a different definition for the duration of reionization which is $\Delta z = (dQ_{\rm HII}/ dz)^{-1}\mid_{Q_{\rm HII}=0.5}$. Using this definition for our models has little effect on the resulting $\Delta z$.}
\label{fig_dz}
\end{figure}

\citet{Monslave17} use single antenna interferometric observations of the cosmological 21-cm signal, to infer a lower limit on the duration of reionization, $\Delta z\gtrapprox 1$. The excluded region is shown as a blue hatched region in Fig.~\ref{fig_dz}. Interestingly, all our models, as well as most other models in the literature \citep[e.g.][]{Robertson15,Mitra15}, are well outside of the excluded region. Therefore, current measurements of the 21~cm brightness temperature do not yet rule out such models.

The duration of reionization can also be constrained using the kinetic Sunyaev Zeldovich (kSZ) effect \citep{George15}. The theory behind this method is well described by \citet{Zahn12}: ionized bubbles form around the first stars and galaxies and grow in size with time and eventually overlap. The motion of these bubbles creates secondary anisotropies in the CMB. The amplitude of the spatially inhomogeneous kSZ power is sensitive to the duration of reionization, $\Delta z$, and, with this method, an upper limit on $\Delta z$ can be placed.  \cite{George15} measured the amplitude of patchiness in the kSZ power spectrum using the {\sc spt} survey, and derived an upper limit on the duration of reionization of $\Delta z\lessapprox 5$; this limit has been recently improved by \cite{Planck16} to $\Delta z\lessapprox 4.8$. The corresponding excluded region
	is the red hatched zone in Fig.~\ref{fig_dz}. Models F10, F10z, F14 and F14z, in which the galaxies on the faint side dominate reionization, are clearly ruled out by these constraints.

\section{Summary and conclusion}
A number of studies suggest that faint galaxies, mostly below the current detection limit of the {\it Hubble Deep Field}, were responsible for reionization \citep[e.g.][]{Ciardi03,Bouwens12,Robertson13}. Such faint galaxies will be challenging to detect, even with the {\em James Webb Space Telescope} \citep{Gardner06}. Here we argued that current data favour a model in which it is the brighter galaxies that dominate. 

Our conclusions are based on computing the rate at which the ionizing emissivity builds-up (Fig.~\ref{fig_E}), contrasting models in which galaxies fainter than UV-magnitude $M_{1500}=-14$ dominate (models F10 and F14), versus models in which galaxies brighter than $M_{1500}=-16$ dominate (model B16). The faint-end of the luminosity function evolves slowly, therefore in models F10 and F14  the reionization process is more extended in redshift as the ionizing emissivity increases only slowly with decreasing redshift, yielding $\Delta z\gtrapprox 8$. In contrast, the bright end of the luminosity function evolves more rapidly, and the reionization process is much less extended in redshift for model B16 $(\Delta z\lessapprox 2$).

Observationally, $\Delta z$ can be constrained by measurements of the global 21~cm brightness temperature \citep{Monslave17} ($\Delta z \gtrapprox 1$) and by the measurements of the amplitude of the patchiness of the kinetic Sunyaev-Zeldovich  effect \citep{George15,Planck16} ($\Delta z \lessapprox 4.8$). These limits clearly disfavour models such as F10 and F14, in which galaxies on the faint-side of the luminosity function dominate. 

Recent theoretical studies \citep[e.g.][]{Sharma15,Sharma17} suggest that galaxies on the bright side of the luminosity function may have produced a greater share of the ionizing emissivity than previously thought. Star forming galaxies at the bright side of the luminosity function are likely to undergo star bursts which drive outflows, thereby facilitating the escape of ionizing photons. This viewpoint is also supported by recent reports of the detection of high escape fraction in local star burst galaxies by \citet{Borthakur14,Izotov16}.

In a scenario such as that represented by our model B16, in which the galaxies on the bright side of the luminosity function dominate the ionizing emissivity, reionization progresses rapidly within a short duration $(\Delta z\leq2)$ (lower panel of Fig.~\ref{fig_Q}), which satisfies the constraints on duration placed by measurements of the 21~cm brightness temperature as well as by the measured amplitude of the patchiness in the kSZ effect (Fig.~\ref{fig_dz}). This suggests that the galaxies on the bright side of the luminosity function (brighter than $M_{1500}=-16$) were the dominant contributors to cosmic reionization. Such galaxies will be easier to study observationally.

 We thank an anonymous referee for their constructive comments. This work was supported by the Science and Technology Facilities Council grant ST/P000541/1.

\footnotesize{\bibliography{ref_faint}}

\begin{thebibliography}{49}
\expandafter\ifx\csname natexlab\endcsname\relax\def\natexlab#1{#1}\fi

\bibitem[{{Borthakur} {et~al}\mbox{.}(2014){Borthakur}, {Heckman}, {Leitherer},
  \& {Overzier}}]{Borthakur14}
{Borthakur} S., {Heckman} T.~M., {Leitherer} C., {Overzier} R.~A., 2014,
  Science, 346, 216

\bibitem[{{Bouwens} {et~al}\mbox{.}(2015{\natexlab{a}}){Bouwens},
  {Illingworth}, {Oesch}, {Caruana}, {Holwerda}, {Smit}, \&
  {Wilkins}}]{Bouwens15}
{Bouwens} R.~J., {Illingworth} G.~D., {Oesch} P.~A., {Caruana} J., {Holwerda}
  B., {Smit} R., {Wilkins} S., 2015{\natexlab{a}}, \apj, 811, 140

\bibitem[{{Bouwens} {et~al}\mbox{.}(2015{\natexlab{b}}){Bouwens},
  {Illingworth}, {Oesch}, {Trenti}, {Labb{\'e}}, {Bradley}, {Carollo}, {van
  Dokkum}, {Gonzalez}, {Holwerda}, {Franx}, {Spitler}, {Smit}, \&
  {Magee}}]{Bouwens14}
{Bouwens} R.~J. {et~al.}, 2015{\natexlab{b}}, \apj, 803, 34

\bibitem[{{Bouwens} {et~al}\mbox{.}(2012){Bouwens}, {Illingworth}, {Oesch},
  {Trenti}, {Labb{\'e}}, {Franx}, {Stiavelli}, {Carollo}, {van Dokkum}, \&
  {Magee}}]{Bouwens12}
{Bouwens} R.~J. {et~al.}, 2012, \apjl, 752, L5

\bibitem[{{Bowman} {et~al}\mbox{.}(2008){Bowman}, {Rogers}, \&
  {Hewitt}}]{Bowman08}
{Bowman} J.~D., {Rogers} A.~E.~E., {Hewitt} J.~N., 2008, \apj, 676, 1

\bibitem[{{Caruana} {et~al}\mbox{.}(2012){Caruana}, {Bunker}, {Wilkins},
  {Stanway}, {Lacy}, {Jarvis}, {Lorenzoni}, \& {Hickey}}]{Caruana12}
{Caruana} J., {Bunker} A.~J., {Wilkins} S.~M., {Stanway} E.~R., {Lacy} M.,
  {Jarvis} M.~J., {Lorenzoni} S., {Hickey} S., 2012, \mnras, 427, 3055

\bibitem[{{Ciardi} {et~al}\mbox{.}(2003){Ciardi}, {Ferrara}, \&
  {White}}]{Ciardi03}
{Ciardi} B., {Ferrara} A., {White} S.~D.~M., 2003, \mnras, 344, L7

\bibitem[{{Faisst}(2016)}]{Faisst16}
{Faisst} A.~L., 2016, ArXiv e-prints : 1605.06507

\bibitem[{{Gardner} {et~al}\mbox{.}(2006){Gardner}, {Mather}, {Clampin},
  {Doyon}, {Greenhouse}, {Hammel}, {Hutchings}, {Jakobsen}, {Lilly}, {Long},
  {Lunine}, {McCaughrean}, {Mountain}, {Nella}, {Rieke}, {Rieke}, {Rix},
  {Smith}, {Sonneborn}, {Stiavelli}, {Stockman}, {Windhorst}, \&
  {Wright}}]{Gardner06}
{Gardner} J.~P. {et~al.}, 2006, \ssr, 123, 485

\bibitem[{{George} {et~al}\mbox{.}(2015){George}, {Reichardt}, {Aird},
  {Benson}, {Bleem}, {Carlstrom}, {Chang}, {Cho}, {Crawford}, {Crites}, {de
  Haan}, {Dobbs}, {Dudley}, {Halverson}, {Harrington}, {Holder}, {Holzapfel},
  {Hou}, {Hrubes}, {Keisler}, {Knox}, {Lee}, {Leitch}, {Lueker}, {Luong-Van},
  {McMahon}, {Mehl}, {Meyer}, {Millea}, {Mocanu}, {Mohr}, {Montroy}, {Padin},
  {Plagge}, {Pryke}, {Ruhl}, {Schaffer}, {Shaw}, {Shirokoff}, {Spieler},
  {Staniszewski}, {Stark}, {Story}, {van Engelen}, {Vanderlinde}, {Vieira},
  {Williamson}, \& {Zahn}}]{George15}
{George} E.~M. {et~al.}, 2015, \apj, 799, 177

\bibitem[{{Gnedin}(2016)}]{Gnedin16}
{Gnedin} N.~Y., 2016, ArXiv e-prints : 1603.07729

\bibitem[{{Haardt} \& {Madau}(2012)}]{Haardt12}
{Haardt} F., {Madau} P., 2012, \apj, 746, 125

\bibitem[{{Izotov} {et~al}\mbox{.}(2016){Izotov}, {Orlitov{\'a}}, {Schaerer},
  {Thuan}, {Verhamme}, {Guseva}, \& {Worseck}}]{Izotov16}
{Izotov} Y.~I., {Orlitov{\'a}} I., {Schaerer} D., {Thuan} T.~X., {Verhamme} A.,
  {Guseva} N.~G., {Worseck} G., 2016, \nat, 529, 178

\bibitem[{{Khaire} {et~al}\mbox{.}(2016){Khaire}, {Srianand}, {Choudhury}, \&
  {Gaikwad}}]{Khaire15}
{Khaire} V., {Srianand} R., {Choudhury} T.~R., {Gaikwad} P., 2016, \mnras, 457,
  4051

\bibitem[{{Kimm} \& {Cen}(2014)}]{Kimm14}
{Kimm} T., {Cen} R., 2014, \apj, 788, 121

\bibitem[{{Loeb} \& {Barkana}(2001)}]{Loeb01}
{Loeb} A., {Barkana} R., 2001, \araa, 39, 19

\bibitem[{{Ma} {et~al}\mbox{.}(2016){Ma}, {Hopkins}, {Kasen}, {Quataert},
  {Faucher-Giguere}, {Keres}, \& {Murray}}]{Ma16}
{Ma} X., {Hopkins} P.~F., {Kasen} D., {Quataert} E., {Faucher-Giguere} C.-A.,
  {Keres} D., {Murray} N., 2016, ArXiv e-prints

\bibitem[{{Madau}(2017)}]{Madau17}
{Madau} P., 2017, ArXiv e-prints

\bibitem[{{Madau} \& {Haardt}(2015)}]{Haardt15}
{Madau} P., {Haardt} F., 2015, \apjl, 813, L8

\bibitem[{{McGreer} {et~al}\mbox{.}(2015){McGreer}, {Mesinger}, \&
  {D'Odorico}}]{McGreer15}
{McGreer} I.~D., {Mesinger} A., {D'Odorico} V., 2015, \mnras, 447, 499

\bibitem[{{McQuinn} {et~al}\mbox{.}(2007){McQuinn}, {Hernquist}, {Zaldarriaga},
  \& {Dutta}}]{Mcquinn07}
{McQuinn} M., {Hernquist} L., {Zaldarriaga} M., {Dutta} S., 2007, \mnras, 381,
  75

\bibitem[{{Mitra} {et~al}\mbox{.}(2015){Mitra}, {Choudhury}, \&
  {Ferrara}}]{Mitra15}
{Mitra} S., {Choudhury} T.~R., {Ferrara} A., 2015, \mnras, 454, L76

\bibitem[{{Mitra} {et~al}\mbox{.}(2016){Mitra}, {Choudhury}, \&
  {Ferrara}}]{Mitra16}
{Mitra} S., {Choudhury} T.~R., {Ferrara} A., 2016, ArXiv e-prints : 1606.02719

\bibitem[{{Monsalve} {et~al}\mbox{.}(2017){Monsalve}, {Rogers}, {Bowman}, \&
  {Mozdzen}}]{Monslave17}
{Monsalve} R.~A., {Rogers} A.~E.~E., {Bowman} J.~D., {Mozdzen} T.~J., 2017,
  \apj, 847, 64

\bibitem[{{Mortlock} {et~al}\mbox{.}(2011){Mortlock}, {Warren}, {Venemans},
  {Patel}, {Hewett}, {McMahon}, {Simpson}, {Theuns}, {Gonz{\'a}les-Solares},
  {Adamson}, {Dye}, {Hambly}, {Hirst}, {Irwin}, {Kuiper}, {Lawrence}, \&
  {R{\"o}ttgering}}]{Mortlock11}
{Mortlock} D.~J. {et~al.}, 2011, \nat, 474, 616

\bibitem[{{Ota} {et~al}\mbox{.}(2008){Ota}, {Iye}, {Kashikawa}, {Shimasaku},
  {Kobayashi}, {Totani}, {Nagashima}, {Morokuma}, {Furusawa}, {Hattori},
  {Matsuda}, {Hashimoto}, \& {Ouchi}}]{Ota08}
{Ota} K. {et~al.}, 2008, \apj, 677, 12

\bibitem[{{Ouchi} {et~al}\mbox{.}(2010){Ouchi}, {Shimasaku}, {Furusawa},
  {Saito}, {Yoshida}, {Akiyama}, {Ono}, {Yamada}, {Ota}, {Kashikawa}, {Iye},
  {Kodama}, {Okamura}, {Simpson}, \& {Yoshida}}]{Ouchi10}
{Ouchi} M. {et~al.}, 2010, \apj, 723, 869

\bibitem[{{Paardekooper} {et~al}\mbox{.}(2011){Paardekooper}, {Pelupessy},
  {Altay}, \& {Kruip}}]{Paardekooper11}
{Paardekooper} J.-P., {Pelupessy} F.~I., {Altay} G., {Kruip} C.~J.~H., 2011,
  \aap, 530, A87

\bibitem[{{Pawlik} {et~al}\mbox{.}(2009){Pawlik}, {Schaye}, \& {van
  Scherpenzeel}}]{Pawlik09}
{Pawlik} A.~H., {Schaye} J., {van Scherpenzeel} E., 2009, \mnras, 394, 1812

\bibitem[{{Planck Collaboration} {et~al}\mbox{.}(2016){Planck Collaboration},
  {Adam}, {Aghanim}, {Ashdown}, {Aumont}, {Baccigalupi}, {Ballardini},
  {Banday}, {Barreiro}, {Bartolo}, {Basak}, {Battye}, {Benabed}, {Bernard},
  {Bersanelli}, {Bielewicz}, {Bock}, {Bonaldi}, {Bonavera}, {Bond}, {Borrill},
  {Bouchet}, {Bucher}, {Burigana}, {Calabrese}, {Cardoso}, {Carron}, {Chiang},
  {Colombo}, {Combet}, {Comis}, {Coulais}, {Crill}, {Curto}, {Cuttaia},
  {Davis}, {de Bernardis}, {de Rosa}, {de Zotti}, {Delabrouille}, {Di
  Valentino}, {Dickinson}, {Diego}, {Dor{\'e}}, {Douspis}, {Ducout}, {Dupac},
  {Elsner}, {En{\ss}lin}, {Eriksen}, {Falgarone}, {Fantaye}, {Finelli},
  {Forastieri}, {Frailis}, {Fraisse}, {Franceschi}, {Frolov}, {Galeotta},
  {Galli}, {Ganga}, {G{\'e}nova-Santos}, {Gerbino}, {Ghosh},
  {Gonz{\'a}lez-Nuevo}, {G{\'o}rski}, {Gruppuso}, {Gudmundsson}, {Hansen},
  {Helou}, {Henrot-Versill{\'e}}, {Herranz}, {Hivon}, {Huang}, {Ili}, {Jaffe},
  {Jones}, {Keih{\"a}nen}, {Keskitalo}, {Kisner}, {Knox}, {Krachmalnicoff},
  {Kunz}, {Kurki-Suonio}, {Lagache}, {L{\"a}hteenm{\"a}ki}, {Lamarre},
  {Langer}, {Lasenby}, {Lattanzi}, {Lawrence}, {Le Jeune}, {Levrier}, {Lewis},
  {Liguori}, {Lilje}, {L{\'o}pez-Caniego}, {Ma}, {Mac{\'{\i}}as-P{\'e}rez},
  {Maggio}, {Mangilli}, {Maris}, {Martin}, {Mart{\'{\i}}nez-Gonz{\'a}lez},
  {Matarrese}, {Mauri}, {McEwen}, {Meinhold}, {Melchiorri}, {Mennella},
  {Migliaccio}, {Miville-Desch{\^e}nes}, {Molinari}, {Moneti}, {Montier},
  {Morgante}, {Moss}, {Naselsky}, {Natoli}, {Oxborrow}, {Pagano}, {Paoletti},
  {Partridge}, {Patanchon}, {Patrizii}, {Perdereau}, {Perotto}, {Pettorino},
  {Piacentini}, {Plaszczynski}, {Polastri}, {Polenta}, {Puget}, {Rachen},
  {Racine}, {Reinecke}, {Remazeilles}, {Renzi}, {Rocha}, {Rossetti}, {Roudier},
  {Rubi{\~n}o-Mart{\'{\i}}n}, {Ruiz-Granados}, {Salvati}, {Sandri},
  {Savelainen}, {Scott}, {Sirri}, {Sunyaev}, {Suur-Uski}, {Tauber}, {Tenti},
  {Toffolatti}, {Tomasi}, {Tristram}, {Trombetti}, {Valiviita}, {Van Tent},
  {Vielva}, {Villa}, {Vittorio}, {Wandelt}, {Wehus}, {White}, {Zacchei}, \&
  {Zonca}}]{Planck16}
{Planck Collaboration} {et~al.}, 2016, ArXiv e-prints : 1605.03507

\bibitem[{{Pritchard} {et~al}\mbox{.}(2015){Pritchard}, {Ichiki}, {Mesinger},
  {Metcalf}, {Pourtsidou}, {Santos}, {Abdalla}, {Chang}, {Chen}, {Weller}, \&
  {Zaroubi}}]{Pritchard15}
{Pritchard} J. {et~al.}, 2015, Advancing Astrophysics with the Square Kilometre
  Array (AASKA14), 12

\bibitem[{{Reed} {et~al}\mbox{.}(2007){Reed}, {Bower}, {Frenk}, {Jenkins}, \&
  {Theuns}}]{Reed07}
{Reed} D.~S., {Bower} R., {Frenk} C.~S., {Jenkins} A., {Theuns} T., 2007,
  \mnras, 374, 2

\bibitem[{{Robertson} {et~al}\mbox{.}(2015){Robertson}, {Ellis}, {Furlanetto},
  \& {Dunlop}}]{Robertson15}
{Robertson} B.~E., {Ellis} R.~S., {Furlanetto} S.~R., {Dunlop} J.~S., 2015,
  \apjl, 802, L19

\bibitem[{{Robertson} {et~al}\mbox{.}(2013){Robertson}, {Furlanetto},
  {Schneider}, {Charlot}, {Ellis}, {Stark}, {McLure}, {Dunlop}, {Koekemoer},
  {Schenker}, {Ouchi}, {Ono}, {Curtis-Lake}, {Rogers}, {Bowler}, \&
  {Cirasuolo}}]{Robertson13}
{Robertson} B.~E. {et~al.}, 2013, \apj, 768, 71

\bibitem[{{Schaerer}(2003)}]{Schaerer03}
{Schaerer} D., 2003, \aap, 397, 527

\bibitem[{{Schenker} {et~al}\mbox{.}(2014){Schenker}, {Ellis}, {Konidaris}, \&
  {Stark}}]{Schenker14}
{Schenker} M.~A., {Ellis} R.~S., {Konidaris} N.~P., {Stark} D.~P., 2014, \apj,
  795, 20

\bibitem[{{Sharma} {et~al}\mbox{.}(2016){Sharma}, {Theuns}, {Frenk}, {Bower},
  {Crain}, {Schaller}, \& {Schaye}}]{Sharma15}
{Sharma} M., {Theuns} T., {Frenk} C., {Bower} R., {Crain} R., {Schaller} M.,
  {Schaye} J., 2016, \mnras, 458, L94

\bibitem[{{Sharma} {et~al}\mbox{.}(2017){Sharma}, {Theuns}, {Frenk}, {Bower},
  {Crain}, {Schaller}, \& {Schaye}}]{Sharma17}
{Sharma} M., {Theuns} T., {Frenk} C., {Bower} R.~G., {Crain} R.~A., {Schaller}
  M., {Schaye} J., 2017, \mnras, 468, 2176

\bibitem[{{Shukla} {et~al}\mbox{.}(2016){Shukla}, {Mellema}, {Iliev}, \&
  {Shapiro}}]{Shukla16}
{Shukla} H., {Mellema} G., {Iliev} I.~T., {Shapiro} P.~R., 2016, \mnras, 458,
  135

\bibitem[{{Singh} {et~al}\mbox{.}(2017){Singh}, {Subrahmanyan}, {Udaya
  Shankar}, {Sathyanarayana Rao}, {Fialkov}, {Cohen}, {Barkana}, {Girish},
  {Raghunathan}, {Somashekar}, \& {Srivani}}]{Singh17}
{Singh} S. {et~al.}, 2017, \apjl, 845, L12

\bibitem[{{Sokasian} {et~al}\mbox{.}(2004){Sokasian}, {Yoshida}, {Abel},
  {Hernquist}, \& {Springel}}]{Sokasian04}
{Sokasian} A., {Yoshida} N., {Abel} T., {Hernquist} L., {Springel} V., 2004,
  \mnras, 350, 47

\bibitem[{{Tilvi} {et~al}\mbox{.}(2014){Tilvi}, {Papovich}, {Finkelstein},
  {Long}, {Song}, {Dickinson}, {Ferguson}, {Koekemoer}, {Giavalisco}, \&
  {Mobasher}}]{Tilvi14}
{Tilvi} V. {et~al.}, 2014, \apj, 794, 5

\bibitem[{{Totani} {et~al}\mbox{.}(2014){Totani}, {Aoki}, {Hattori}, {Kosugi},
  {Niino}, {Hashimoto}, {Kawai}, {Ohta}, {Sakamoto}, \& {Yamada}}]{Totani14}
{Totani} T. {et~al.}, 2014, \pasj, 66, 63

\bibitem[{{Trebitsch} {et~al}\mbox{.}(2017){Trebitsch}, {Blaizot}, {Rosdahl},
  {Devriendt}, \& {Slyz}}]{Trebitsch17}
{Trebitsch} M., {Blaizot} J., {Rosdahl} J., {Devriendt} J., {Slyz} A., 2017,
  \mnras, 470, 224

\bibitem[{{van Haarlem} {et~al}\mbox{.}(2013){van Haarlem}, {Wise}, {Gunst},
  {Heald}, {McKean}, {Hessels}, {de Bruyn}, {Nijboer}, {Swinbank}, {Fallows},
  {Brentjens}, {Nelles}, {Beck}, {Falcke}, {Fender}, {H{\"o}randel},
  {Koopmans}, {Mann}, {Miley}, {R{\"o}ttgering}, {Stappers}, {Wijers},
  {Zaroubi}, {van den Akker}, {Alexov}, {Anderson}, {Anderson}, {van Ardenne},
  {Arts}, {Asgekar}, {Avruch}, {Batejat}, {B{\"a}hren}, {Bell}, {Bell}, {van
  Bemmel}, {Bennema}, {Bentum}, {Bernardi}, {Best}, {B{\^i}rzan}, {Bonafede},
  {Boonstra}, {Braun}, {Bregman}, {Breitling}, {van de Brink}, {Broderick},
  {Broekema}, {Brouw}, {Br{\"u}ggen}, {Butcher}, {van Cappellen}, {Ciardi},
  {Coenen}, {Conway}, {Coolen}, {Corstanje}, {Damstra}, {Davies}, {Deller},
  {Dettmar}, {van Diepen}, {Dijkstra}, {Donker}, {Doorduin}, {Dromer}, {Drost},
  {van Duin}, {Eisl{\"o}ffel}, {van Enst}, {Ferrari}, {Frieswijk}, {Gankema},
  {Garrett}, {de Gasperin}, {Gerbers}, {de Geus}, {Grie{\ss}meier}, {Grit},
  {Gruppen}, {Hamaker}, {Hassall}, {Hoeft}, {Holties}, {Horneffer}, {van der
  Horst}, {van Houwelingen}, {Huijgen}, {Iacobelli}, {Intema}, {Jackson},
  {Jelic}, {de Jong}, {Juette}, {Kant}, {Karastergiou}, {Koers}, {Kollen},
  {Kondratiev}, {Kooistra}, {Koopman}, {Koster}, {Kuniyoshi}, {Kramer},
  {Kuper}, {Lambropoulos}, {Law}, {van Leeuwen}, {Lemaitre}, {Loose}, {Maat},
  {Macario}, {Markoff}, {Masters}, {McFadden}, {McKay-Bukowski}, {Meijering},
  {Meulman}, {Mevius}, {Middelberg}, {Millenaar}, {Miller-Jones}, {Mohan},
  {Mol}, {Morawietz}, {Morganti}, {Mulcahy}, {Mulder}, {Munk}, {Nieuwenhuis},
  {van Nieuwpoort}, {Noordam}, {Norden}, {Noutsos}, {Offringa}, {Olofsson},
  {Omar}, {Orr{\'u}}, {Overeem}, {Paas}, {Pandey-Pommier}, {Pandey}, {Pizzo},
  {Polatidis}, {Rafferty}, {Rawlings}, {Reich}, {de Reijer}, {Reitsma},
  {Renting}, {Riemers}, {Rol}, {Romein}, {Roosjen}, {Ruiter}, {Scaife}, {van
  der Schaaf}, {Scheers}, {Schellart}, {Schoenmakers}, {Schoonderbeek},
  {Serylak}, {Shulevski}, {Sluman}, {Smirnov}, {Sobey}, {Spreeuw}, {Steinmetz},
  {Sterks}, {Stiepel}, {Stuurwold}, {Tagger}, {Tang}, {Tasse}, {Thomas},
  {Thoudam}, {Toribio}, {van der Tol}, {Usov}, {van Veelen}, {van der Veen},
  {ter Veen}, {Verbiest}, {Vermeulen}, {Vermaas}, {Vocks}, {Vogt}, {de Vos},
  {van der Wal}, {van Weeren}, {Weggemans}, {Weltevrede}, {White}, {Wijnholds},
  {Wilhelmsson}, {Wucknitz}, {Yatawatta}, {Zarka}, {Zensus}, \& {van
  Zwieten}}]{van-Haarlem13}
{van Haarlem} M.~P. {et~al.}, 2013, \aap, 556, A2

\bibitem[{{Wise} \& {Cen}(2009)}]{Wise09}
{Wise} J.~H., {Cen} R., 2009, \apj, 693, 984

\bibitem[{{Wise} {et~al}\mbox{.}(2014){Wise}, {Demchenko}, {Halicek}, {Norman},
  {Turk}, {Abel}, \& {Smith}}]{Wise14}
{Wise} J.~H., {Demchenko} V.~G., {Halicek} M.~T., {Norman} M.~L., {Turk} M.~J.,
  {Abel} T., {Smith} B.~D., 2014, \mnras, 442, 2560

\bibitem[{{Yajima} {et~al}\mbox{.}(2011){Yajima}, {Choi}, \&
  {Nagamine}}]{Yajima11}
{Yajima} H., {Choi} J.-H., {Nagamine} K., 2011, \mnras, 412, 411

\bibitem[{{Zahn} {et~al}\mbox{.}(2012){Zahn}, {Reichardt}, {Shaw}, {Lidz},
  {Aird}, {Benson}, {Bleem}, {Carlstrom}, {Chang}, {Cho}, {Crawford}, {Crites},
  {de Haan}, {Dobbs}, {Dor{\'e}}, {Dudley}, {George}, {Halverson}, {Holder},
  {Holzapfel}, {Hoover}, {Hou}, {Hrubes}, {Joy}, {Keisler}, {Knox}, {Lee},
  {Leitch}, {Lueker}, {Luong-Van}, {McMahon}, {Mehl}, {Meyer}, {Millea},
  {Mohr}, {Montroy}, {Natoli}, {Padin}, {Plagge}, {Pryke}, {Ruhl}, {Schaffer},
  {Shirokoff}, {Spieler}, {Staniszewski}, {Stark}, {Story}, {van Engelen},
  {Vanderlinde}, {Vieira}, \& {Williamson}}]{Zahn12}
{Zahn} O. {et~al.}, 2012, \apj, 756, 65

\end{thebibliography}

\end{document}